# Plasmonic random laser on an optical fiber tip

Dipendra S. Khatri,[1] Ying Li,[2] Jiyang Chen,[1] Anna Elizabeth Stocks,[1] Elyahb Allie Kwizera,[3] Xiaohua Huang,[3] Christos Argyropoulos,[2,†] and Thang Hoang[1,*]

[1]Department of Physics and Material Science, The University of Memphis, Memphis, TN 38152

[2]Department of Electrical and Computer Engineering, University of Nebraska-Lincoln, Lincoln, NE 68588

[3]Department of Chemistry, The University of Memphis, Memphis, TN 38152

[†]christos.argyropoulos@unl.edu
[*]tbhoang@memphis.edu

Random lasing occurs as the result of a coherent optical feedback from multiple scattering centers. Here, we demonstrate that plasmonic gold nanostars are efficient light scattering centers, exhibiting strong field enhancement at their nanotips, which assists a very narrow bandwidth and highly amplified coherent random lasing with a low lasing threshold. First, by embedding plasmonic gold nanostars in a rhodamine 6G dye gain medium, we observe a series of very narrow random lasing peaks with full-width at half-maximum ~ 0.8 nm. In contrast, free rhodamine 6G dye molecules exhibit only a single amplified spontaneous emission peak with a broader linewidth of 6 nm. The lasing threshold for the dye with gold nanostars is two times lower than that for a free dye. Furthermore, by coating the tip of a single-mode optical fiber with gold nanostars, we demonstrate a collection of random lasing signal through the fiber that can be easily guided and analyzed. Time-resolved measurements show a significant increase in the emission rate above the lasing threshold, indicating a stimulated emission process. Our study provides a method for generating random lasing in the nanoscale with low threshold values that can be easily collected and guided, which promise a range of potential applications in remote sensing, information processing, and on-chip coherent light sources.

The recently developed nanolasers rely on the stimulated emission of gain media in optical cavities [1-4]. Such stimulated emission processes result in very well-determined lasing characteristics including spectral position, bandwidth, and pumping threshold. Significant advances in fabrication techniques have also recently enabled a new class of "mirrorless" plasmon assisted coherent lasing action from ordered metallic nanoparticle lattices [5-7]. Similarly for the latter case, the lasing characteristics are also very well determined by a set of predefined parameters such as the nanoparticle's size, lattice or dielectric constant. Generally, periodic lattices of plasmonic nanoparticles provide well-determined surface lattice plasmon resonances which trigger lasing action in an embedded gain medium. Interestingly, it has been demonstrated that a random distribution of plasmonic nanostructures could also lead to amplified spontaneous emission and coherent random lasing due to multiple scattering processes [8-10]. It has been demonstrated that the surface plasmon resonances of metallic nanoparticles can act as scattering centers and subsequently form random closed-loop cavities to efficiently boost the lasing process [8, 10-15]. Furthermore, the coherent random lasing has previously been observed in many different disordered materials and such observations are sometimes related to the Anderson light localization, an analogous phenomenon to defect centers in electronic systems [16-18].

Plasmon-assisted amplified spontaneous emission and coherent random lasing by using various nanostructured plasmonic platforms and gain media have been recently demonstrated. Gold (Au) or silver (Ag) nanorods [11], nanospheres [14], nanoflowers [13] and nanostars [8, 9] can act as efficient scattering centers. Plasmonic metasurfaces [15, 19] and channel waveguides [10, 12] have also been demonstrated to support random lasing. Among various plasmonic nanostructures, Au nanostars [8, 9, 13] have been shown to be a very effective mean to boost the lasing efficiency over spectral bandwidth. More specifically, previous studies [8, 9] have shown that Au nanostars, through their nanoscale tips, are very efficient scattering centers over a very broad spectral range, which can be used to scatter the light emission from various dye emitters. In this work, we experimentally and theoretically demonstrate the efficient random lasing process by using core-shell iron oxide-Au nanostars. The rhodamine 6g (R6G) dye is used as a gain medium. We demonstrate a two-fold reduction to the pumping threshold and much narrower spectral lasing linewidth in the case of dye with the presence of nanostars compared to the dye without nanostars. Furthermore, by coating the tip of a single mode optical fiber with a layer of Au nanostars, we demonstrate the collection of the random lasing emission directly into the fiber. This allows us to easily out couple the random lasing signal, which often suffers a significant loss from the spatial incoherence, to external optics and detection instruments.

The sample structure of this present study consists of colloidal synthesized iron oxide-gold core-shell nanostars with an approximate size of ~ 75 nm (Fig. 1(a)) and R6G dye gain medium. The magnetic iron oxide core is approximately 35 nm and the Au shell layer is 40 nm. The original purpose of iron oxide-gold core-shell nanostars was to make use of its dual-functionality, namely magnetic and plasmonic properties, for biomedical applications [20]. However, for this particular work, the iron core's role is to serve as a catalyst for the nanostar growth and its influence on the plasmon resonances of the Au nanostar is negligible. The spikes (or tips) of the Au nanostars are estimated, via Transmission Electron Microscopy (TEM) image (Fig. 1(b)), to be less than 10 nm. Indeed, the small tips of the nanostar are permitting the formation of extremely tiny nanoscale hot spots as schematically shown in Fig. 1(c).

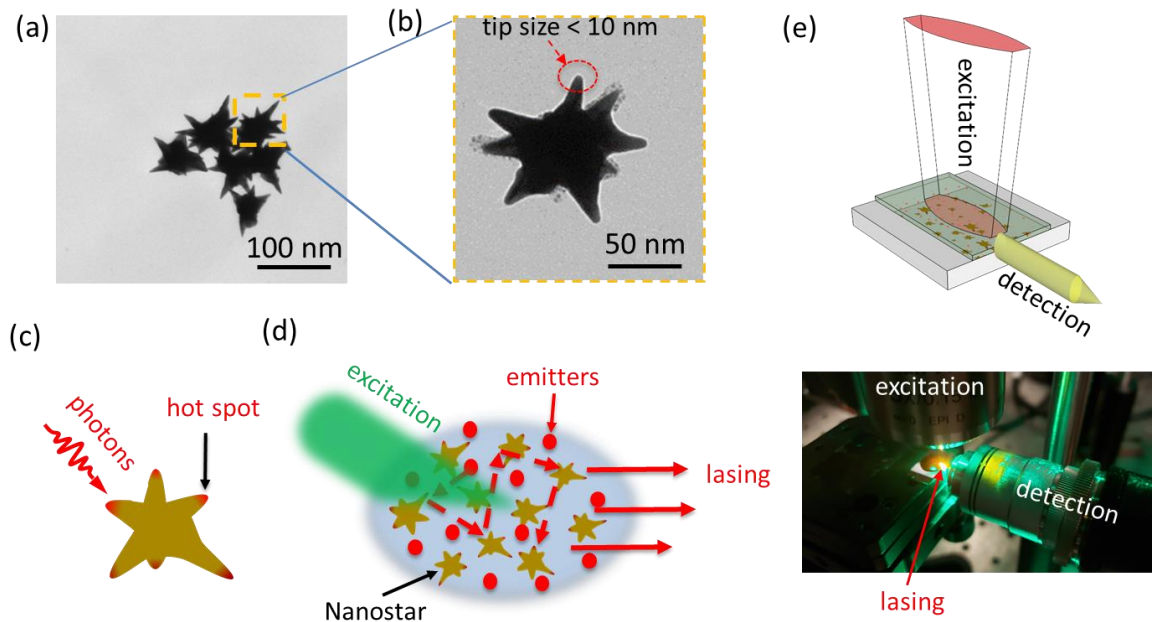

**Fig. 1.** Plasmon assisted random lasing by using gold nanostars. (a)-(b) TEM images of the Au nanostars. (c) Schematic of the scattering of photons and the formed hot spots at the tips of the plasmonic nanostar. (d) Schematic of lasing as a result of an amplification due to random scattering events. (e) Top: schematic of excitation and detection in an analogous to a waveguide configuration. Bottom: a picture of the actual experimental setup - the lasing signal appears in yellow.

In the experimental procedure, Au nanostars with concentration 20 mM are mixed with a R6G dye solution at concentration 0.01 mM in dimethyl sulfoxide, as schematically shown in Fig. 1(d).

These concentrations are similar to parameters previously determined by Ziegler et al [8]. We have tried various concentrations and observed that for dye concentrations less than 0.005 mM, the lasing did not occur. The mixed nanostars/dye suspension is excited by a short-pulsed solid-state laser at 515 nm (Coherent Flare NX, pulse length shorter than 1.0 ns, maximal pulse energy 322 µJ, 2 kHz repetition date). For efficient excitation and collection, the laser excitation beam was shaped into a thin stripe (~ 1 mm) as shown in Fig. 1(e) (top) by using a pair of cylindrical lenses. The excitation laser is from the top of the sample and the signal collection is from a side facet, in an analogous set-up to a waveguide configuration. The pump laser is filtered by a 550 nm long-pass filters (Semrock). The lower panel of Fig. 1(e) shows an actual picture of the setup. The collected lasing signal is fed into a multiple mode fiber and analyzed by a portable spectrometer (Ocean optics USB2000+). Furthermore, to integrate the random lasing structure with a single mode fiber, the fiber's tip is coated with a layer of Au nanostars and then the coated tip is inserted directly into a free R6G dye solution (without nanostar suspension). In this latter case, the lasing signal is collected directly via the single mode fiber and the detection objective lens shown in the lower panel of Fig. 1(e) is not needed. The R6G dye and nanostar absorption measurements are carried out by using a Cary 100 UV-Vis spectrometer. For the time-resolved single photon counting measurements, the lasing signal from a single mode fiber is filtered by another spectrometer (Horiba iHR550), guided through a second exit port and collected by a fast-timing avalanche photodiode. A time-correlated single photon counting module (PicoHarp 300) with a time bin of 16 ps is used to analyze the number of photons as a function of time when they arrived at the photodiode. Final computed lifetimes are obtained from fits to the data deconvolved with the instrument's response function [21].

Figure 2 shows the absorption and emission spectra of the R6G dye in dimethyl sulfoxide solution along with the absorption response of the Au nanostars. Both the absorption and emission spectra of the R6G dye have significant overlaps with the broad resonance of the nanostars from 480 nm to 800 nm. The plasmonic resonance of the Au nanostars plays an important role in boosting the absorption rate of the dye and increases the scattering efficiency, which leads to an increase in the number of the dye's emitted photons. Moreover, the laser excitation wavelength is choosen to be 515 nm, which is very close to the absorption peak of the R6G dye (~ 520 nm), hence, helping to further increase the dye absorption rate.

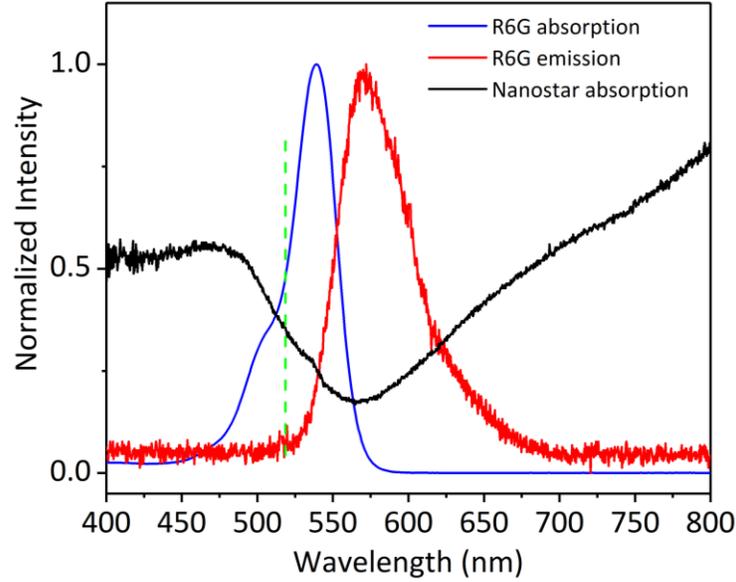

**Fig. 2.** The R6G dye absorption (blue), emission (red), and nanostar absorption (black) curves. The laser excitation wavelength (515 nm) is shown as the dashed green line.

Important characteristics of a lasing system include the existence of a pumping threshold, narrowing of the emission spectrum, shortening of the decay time, and most importantly the resulted coherent light emission. The top left panel of Fig. 3 displays the emission spectra of a free R6G dye solution (without nanostars) at various excitation powers, which exhibit a clear threshold at an excitation density of $P_{th} = 0.16 \; mJ/cm^2$ (measured before the excitation objective lens). At low excitation densities, the dye molecule emission is characterized by a broad spontaneous emission spectrum. As the excitation density increases, a narrow, stimulated emission peak appears at ~580 nm with a full-width at half-maximum (FWHM) of 6 nm. It is observed that, in Au nanostars suspension with the same dye concentration, the lasing behavior occurs at a lower threshold ($0.078 \; mJ/cm^2$), as shown in Fig. 3 (c). Furthermore, with the presence of Au nanostars, the above-threshold lasing signal emerges as a series of very narrow peaks with FWHM ~ 0.8 nm. We would like to point out that due to the random distribution of Au nanostars in the solution, the closed-loop cavities formed by these nanostars are also random. Therefore, the narrow peaks appear to be random from one measurement to another, even with a very short 100 ms integration time. The result of our random lasing measurements with Au nanostars is in good agreement with previous observations [8, 9]. Indeed, the plasmon assisted random lasing peaks observed in our study have a narrower FWHM compared with lasing peaks from free dye or even in the case of lasing action observed by periodic lattices of plasmonic nanoparticles [5, 6].

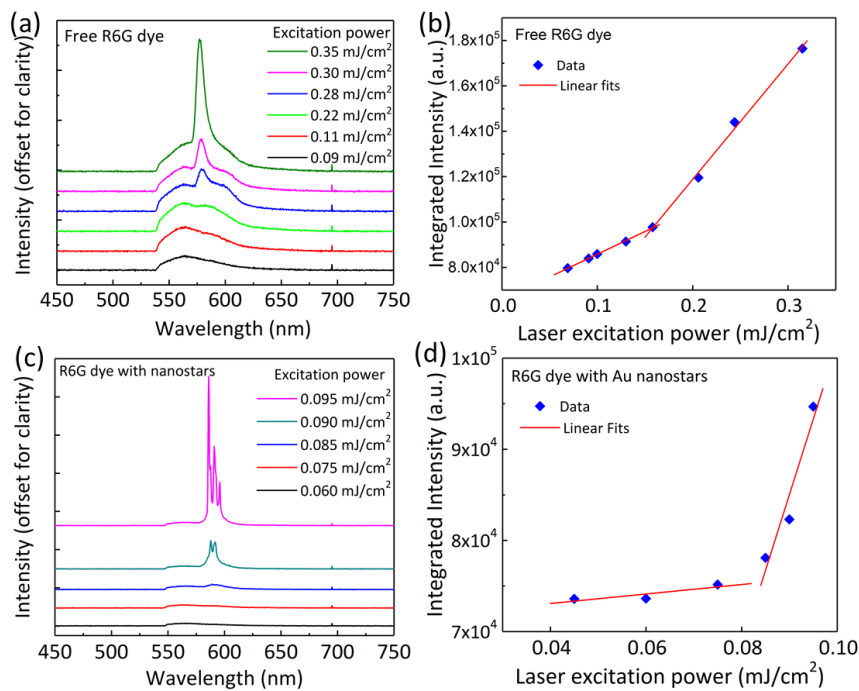

**Fig. 3.** Random lasing by nanostars. (a) Emission spectra of the free R6G dye solution at different excitation powers (100 ms acquisition time). (b) Integrated emission intensity as a function of the excitation power for free R6G dye. (c) Emission spectra of the R6G dye/nanostar suspension at different excitation powers. (d) Integrated emission intensity as a function of the excitation power for the R6G dye/nanostar suspension.

One of the limitations of lasers with mirrorless feedback is a lack of spatial coherence [16, 22, 23]. This can potentially raise an issue for applications relevant to long distance propagation or applications that need directional light emission. Attempts to combine the random lasing with waveguides have been demonstrated [10, 12, 24]. However, these approaches still face the problem on how to couple the random lasing signal to external optics and other detection instruments. Abaie et al [17] have also investigated the random lasing in an Anderson localizing optical fiber. To overcome this issue, we demonstrate the random lasing process on the tip of an optical fiber which can directly collect and guide the optical mode. In order to demonstrate this we integrate Au nanostars, the scattering centers where the random lasing occurs, on the tip of a bare optical fiber. The coated fiber tip is inserted into the R6G dye as schematically shown in Fig. 4(a). To ensure that the Au nanostars adhere to the fiber tip we use the following procedure: first, the Au nanostars are suspended in a polyethylene glycol electrolyte solution which resulted in a slightly negative surface charge for the nanoparticles. Second, the tip of the bare optical fiber is dip coated with a very thin layer (~ 1 nm) of poly(allylamine hydrochloride) polymer, which is slightly positive. The optical fiber tip is subsequently dipped into the Au nanostar solution for 30 minutes to allow for the nanostars to adhere to the fiber tip via the Coulomb attraction force. Figures 4(b)-(c) show scanning electron microscope images of the Au nanostars at the core of the cleaved facet and the outer side of the fiber tip.

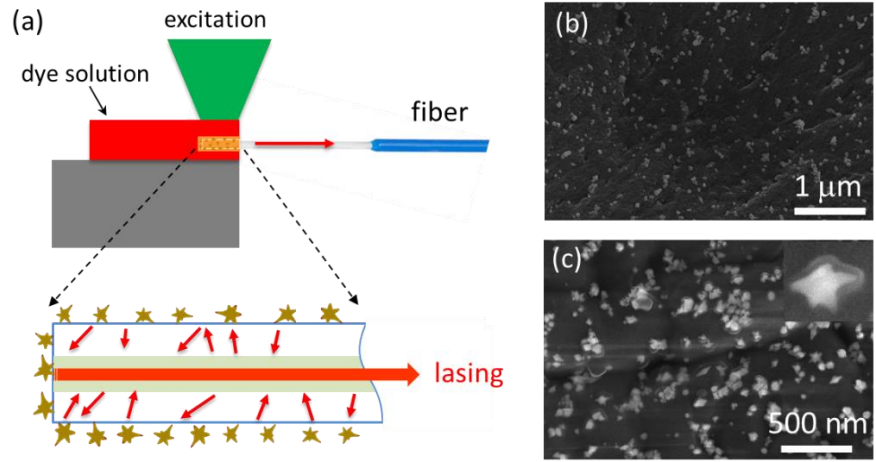

**Fig. 4.** Nanostar integrated single mode fiber for random lasing. (a) Schematic of the nanostar-coated bare single mode fiber tip inserted in the R6G dye solution. (b)-(c) SEM images of the nanostars at the end facet and on the side of the bare single mode fiber, respectively. The inset in (c) shows a zoom-in image of a single nanostar.

Figure 5 shows the results of measurements for two different cases: from a bare fiber tip without Au nanostars (left) and a bare fiber tip coated with Au nanostars (right) inserted into a free R6G dye solution. For both cases, the other end of the fiber is connected to a fiber-coupled spectrometer. It is obvious that for a bare fiber without the Au nanostars, we observed a similar broad lasing characteristic to the free dye. It is notable that the laser excitation power in this case is higher compared with the measurements shown in Fig. 3(a) due to the limited numerical aperture (NA) of the single mode fiber (NA = 0.14) and lack of scattering centers. For the second case, i.e. nanostar-coated fiber, we observe very narrow peaks at the above-threshold laser excitation power. Similar to the case where nanostars are suspended in the dye solution, we observe strong narrow peaks with the lowest FWHM of 0.6 nm. Because the nanostars are coulombically attracted to the fiber's tip, their positions were fixed during the measurements. Therefore, in contrast to the solution suspension sample where nanostars could move freely, a closed-loop cavity at the tip of the fiber is fixed and the random lasing emission wavelength is stable. Furthermore, the nanostars at the end facet (Fig. 4(a)) and the side of the fiber tip (Fig. 4(c)) acted as efficient scattering centers which scatter the lasing photons into the fiber's core at random angles thereby increasing the output signal's intensity.

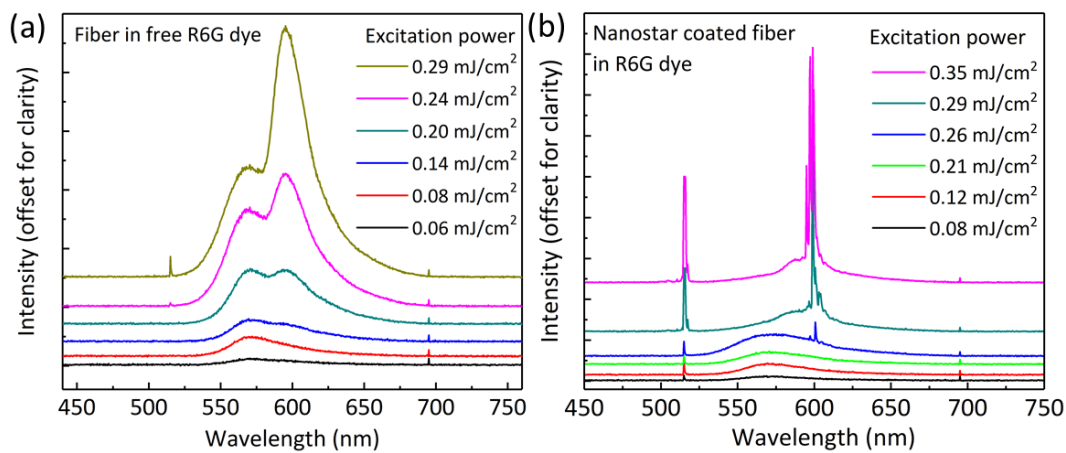

**Fig. 5.** Lasing through the tip of an optical fiber. (a) Emission spectra collected from a single mode fiber inserted into a dye solution at various laser excitation powers. (b) Similar as (a) but the fiber tip is coated with a layer of Au nanostars. The sharp spectral feature showed up at 515 nm in both (a) and (b) is the excitation laser which scattered into the fiber.

One of the signatures of the stimulated emission is a reduction in the decay time of the emitted photons. At the lasing emission wavelength of 600 nm (Fig. 5(b)), time-resolved measurements of the emitted photons showed a reduction of the decay time when the excitation power density increases from below to above the lasing threshold (from ~ 1 ns to < 100 ps), indicating a transition from the spontaneous emission to the stimulated emission regime. Indeed, the measured decay time of the lasing signal above the threshold is limited by the pulse width of the excitation laser and the true lifetime could be much shorter.

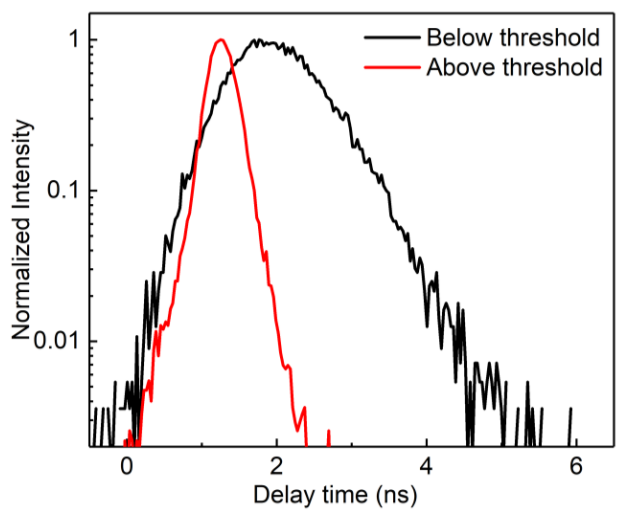

**Fig. 6.** Time-resolved measurements of the dye emission from the nanostar-coated fiber tip for excitation power below (black) and above (red) the lasing threshold.

Next, we perform a numerical study of the plasmon assisted random lasing and evaluate the lasing behavior by using full-wave simulations (COMSOL Multiphysics software). A previous computational study by Solis et al [25] has shown that several nanoscale hotspots of confined light exist in an individual nanostar which cause the nanostars with longer tips to produce red-shifted plasmonic resonances. In addition to the field enhancement (hotspots) around a single nanostar, strong plasmon coupling can also occur when different neighboring nanostars approach each other and form a nanocluster which leads to additional hotspots that are beneficial for achieving random lasing. This is shown in the right inset of Fig. 7, where plasmon coupling and hotspot formations are predicted through full-wave simulations when two nanostar tips approach each other and form a simple nanocluster. For simplicity, and to reduce the modeling computational burden, we only use two dissimilar nanostars in the cluster, each of them containing 12 equal-length uniformly distributed sharp tips branching out from a central spherical core. The dimensions of nanostars are predicted from the actual shapes of the particles taken from the TEM images (Fig. 1(a)) with the same core diameter (~50 nm) and slightly different tip lengths ($NS_1$ ~30 nm, $NS_2$ ~20 nm). Smaller separation distances lead to higher field intensity enhancement at the formed hotspots, and, the center to center distance between the two nanostars ($NS_1$, $NS_2$) is set to be equal to 90 nm. The dispersive dielectric properties of Au are taken by previously derived experimental data [26]. The surrounding space is loaded with R6G dye, i.e., the used active (gain) medium with a relative permittivity at steady state equal to $\varepsilon = \varepsilon_r - i\delta$, where the real part is set to $\varepsilon_r = 1.82$ to match the resonances appearing in the experimental results and the imaginary part corresponds to the gain coefficient that is set to be an arbitrary variable with constant value $\delta$. This is a valid approximation to model active materials at the steady-state condition [27-29], which can be implemented in full-wave simulations. The nanoclusters composed of metal nanostars have random orientations, as seen in Figs. 4(a) and 4(c). Therefore, the extinction spectrum is simulated and computed by averaging over all possible orientations of the complex nanocluster, which is equivalent to fixing the internal axes of the nanocluster and average over all incident wave directions [30]. Figure 7 presents the simulated extinction cross section (ECS) spectra for the nanocluster of two nanostars (shown in the right side of Fig. 7) suspended in the active medium (R6G dye) by using different small values of the gain coefficient $\delta$, where the active system is normally illuminated under an unpolarized plane wave. In our simulations, the increase in the gain coefficient $\delta$ directly corresponds to the rise in the excitation power demonstrated before through the experimental results. Note that a random lasing response is clearly observed at $\delta = 0.33$, where a giant and narrow ECS peak combined with several other smaller peaks exist not only due to the hotspots formation around the individual nanotips but also due to the large plasmonic coupling between adjacent nanostars. The simulation results match very well the experimentally obtained results shown before in Fig. 3(c). The associated normalized field distribution (*xz*-plane cross section) is shown in the bottom right inset of Fig. 7 under a *x*-polarized plane wave illumination. It is clear that a large field enhancement, reaching to 400, can be achieved due to both plasmonic coupling and hotspot area formations, which are proven to be the main mechanisms of the experimentally realized random lasing process.

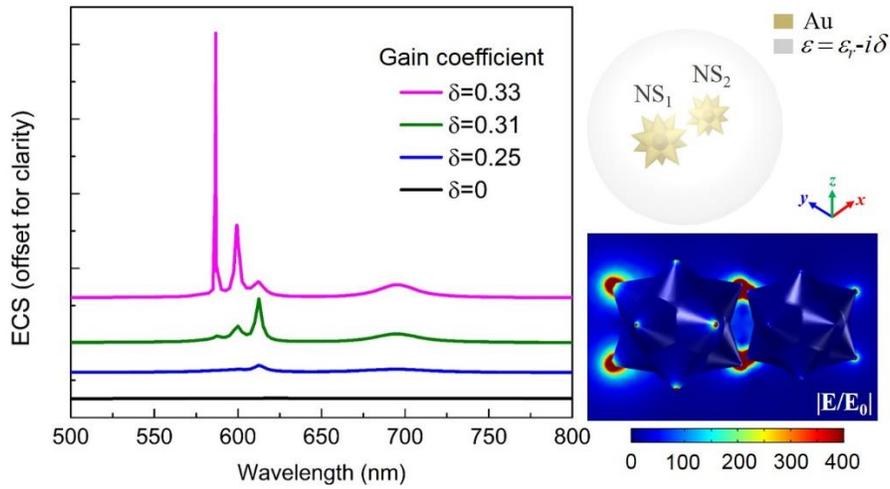

**Fig. 7**. Simulated extinction spectra of a nanocluster of two nanostars suspended in the active medium (R6G dye) varying with different small values of the gain coefficient $\delta$. The top right inset is the geometry of the simulated system consisting of two Au nanostars inside a spherical domain loaded with the active dielectric material. Bottom right inset: normalized field distribution of two coupled nanostars. The gain coefficient is fixed to $\delta = 0.33$ and the incidence wavelength is 600 nm.

In conclusion, we have demonstrated the random lasing by the scattering of emitted photons from plasmonic Au nanostars embedded in a R6G dye medium. The highly engineered sub-10 nm tips of the nanostars provide effective scattering centers. The resulted randomized closed-loop cavities led to sub-nm lasing bandwidth and low threshold pumping power. We have also demonstrated a method to directly guide the random lasing mode into a single mode optical fiber which provides a convenient means to collect and guide the poorly spatial coherent random lasing through guided modes. The obtained experimental results were modeled by full-wave simulations which prove that both plasmonic coupling and hotspot area formations are the main mechanisms of the realized random lasing process. Our study provides a platform for a range of potential applications in remote sensing, information processing, and on-chip coherent nanoscale light sources.

ACKNOWLEDGMENTS

This work is supported by the National Science Foundation (NSF) (Grant # DMR-1709612). TH acknowledges the Ralph E. Powe Junior Faculty Enhancement Award from Oak Ridge Associated Universities. XH acknowledges the support from the National Institutes of Health (Grant No. 1R15 CA 195509-01). TH and XH acknowledge the support from the FedEx Institute of Technology at the University of Memphis.